\documentclass[letterpaper,english,preprint,aps,prl,footinbib,amssymb,floatfix,superscriptaddress]{revtex4-1}
\usepackage[T1]{fontenc}
\usepackage[latin9]{inputenc}
\usepackage{textcomp}
\usepackage{amsmath}
\usepackage{graphicx}

\makeatletter

\usepackage{babel}

\makeatother

\begin{document}

\title{Snell's Law for Spin Waves}

\author{J. Stigloher}

\author{M. Decker}

\author{H.S. K{\"o}rner}

\address{Department of Physics, Regensburg University, 93053 Regensburg, Germany}

\author{K. Tanabe}

\address{Department of Physics, Nagoya University, Nagoya, Aichi 464-8602,
Japan}

\author{T. Moriyama}

\author{T. Taniguchi}

\author{H. Hata}

\address{Institute for Chemical Research, Kyoto University, Uji, Kyoto 611-0011,
Japan}

\author{M. Madami}

\address{Dipartimento di Fisica e Geologia, Universita di Perugia, I-06123
Perugia, Italy}

\author{G. Gubbiotti}

\address{Istituto Officina dei Materiali del Consiglio Nazionale delle Ricerche (IOM-CNR), Sede di Perugia, c/o Dipartimento di Fisica e Geologia, Via A. Pascoli, I-06123 Perugia, Italy}

\author{K. Kobayashi}

\address{Department of Physics, Osaka University, Toyonaka, Osaka 560-0043,
Japan}

\author{T. Ono}

\address{Institute for Chemical Research, Kyoto University, Uji, Kyoto 611-0011,
Japan}

\author{C.H. Back}

\address{Department of Physics, Regensburg University, 93053 Regensburg, Germany}

\date{\today}

\begin{abstract}
We report the experimental observation of Snell's law for magneto-static spin waves in thin ferromagnetic Permalloy films by imaging incident, refracted and reflected waves. We use a thickness step as the interface between two media with different dispersion relation. Since the dispersion relation for magneto-static waves in thin ferromagnetic films is anisotropic, deviations from the isotropic Snell's law known in optics are observed for incidence angles larger than 25\textdegree{} with respect to the interface normal between the two magnetic media. Furthermore, we can show that the thickness step modifies the wavelength and the amplitude of the incident waves. Our findings open up a new way of spin wave steering for magnonic applications.
\end{abstract}
\maketitle

Snell's law describes the refraction of waves at the transition between two media with different indices of refraction. In optics the dispersion relation of light is isotropic and thus the relation between the incident and refracted angles is solely determined by the ratio of the refractive indices. In contrast, for spin waves in thin films with in-plane magnetization the dispersion relation is inherently anisotropic and thus deviations from Snell's law in optics are expected \cite{Gruszecki2015,Dadoenkova2012,Yasumoto1983,Gieniusz2014,Gorobets1998,Jeong2011} but have so far not been reported directly. 

In the emerging field of magnonics, it is foreseen that spin waves can be used as carriers transmitting information from one medium to another. Thus, it is important to study their refraction and reflection at the interface between two magnetic media. Furthermore, their efficient manipulation and steering is one of the fundamental problems that needs to be solved before spin waves or magnons can be used in magnonic devices \cite{Kruglyak2010,Vogt2014}. Attempts in steering range from using artificially designed magnonic crystals \cite{Chumak2010,Duerr2011,Haldar2016} to spin wave guiding in nanostructures \cite{Vogt2012,Wagner2016}.

Most magnonic devices realized so far are rather large \cite{Chumak2015}, although they could potentially be scaled down to the nanometer range. One reason is that spin waves are typically generated by lithographically defined microwave antennas that limit the experimentally accessible wavelengths to a few hundred nanometers. There is major interest in overcoming this limit and different schemes have been proposed \cite{Yu2013,Demidov2011}. 

\begin{figure*}
	\includegraphics[width=1\textwidth]{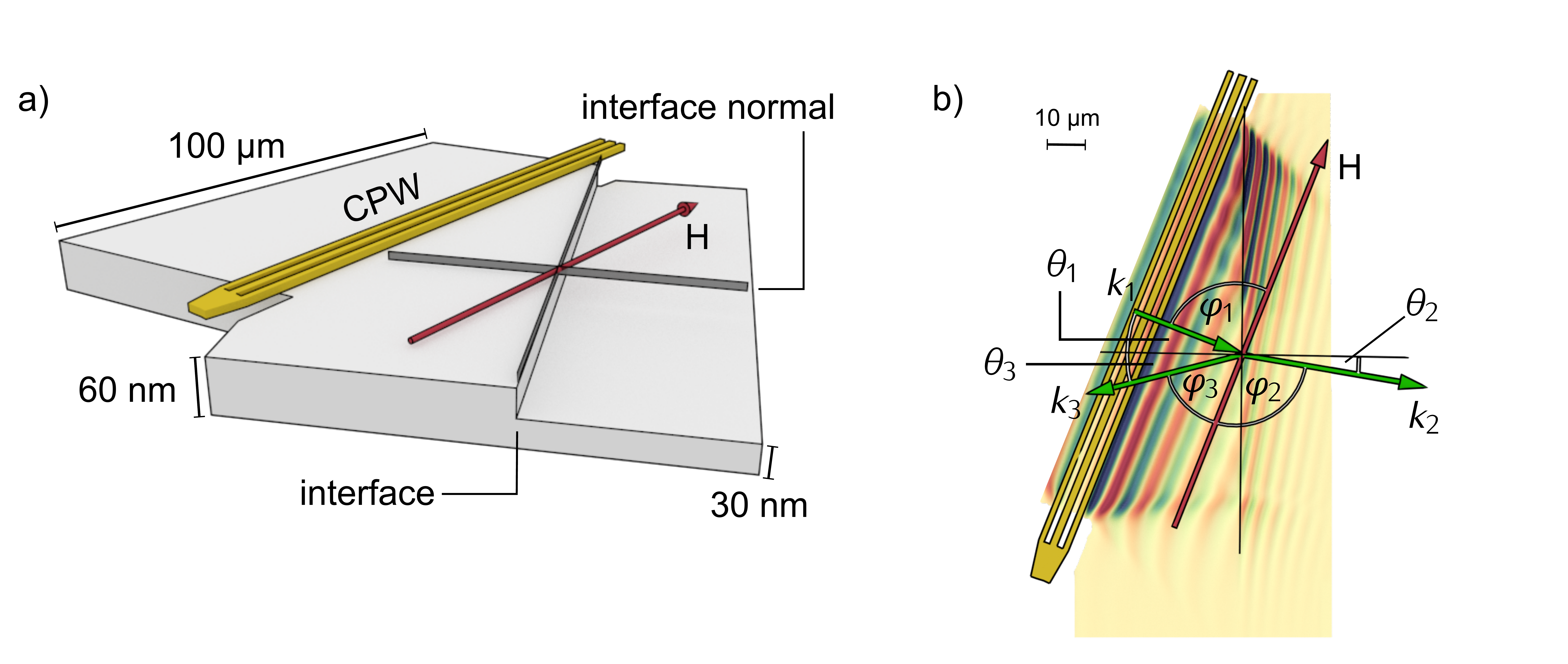} 
	\caption{\label{fig:Sketch-of-the-experiment}  a) Sketch
		of the sample with the $\it{z}$-axis not drawn to scale.
		The red arrow indicates the direction of the externally applied magnetic field which is 
		aligned parallel to the coplanar wave guide (yellow). The latter is used to
		excite spin waves which propagate perpendicular to it. b) Top view of a) with exemplary data acquired by TRMOKE. The green arrows show the wave vectors  $k_1$, $k_2$ and $k_3$ relevant for the analysis.  $\varphi_{1-3}$ denote the angles of the wave vectors with respect to the external field, while $\theta_{1-3}$ denote the angles with respect to the interface normal. The indices 1--3 correspond to the incident, refracted and reflected wave, respectively.}
\end{figure*}

In the experiments presented here, we use Snell's law for spin waves in the dipolar regime as an efficient means of spin wave steering and as a way to reach lower wavelengths. We use a thickness step to realize the transition between two magnetic media with different dispersion relations for propagating spin waves. Spin waves are excited in a thick Permalloy film and subsequently propagate into a film with lower thickness, see Fig. \ref{fig:Sketch-of-the-experiment} a). This idea \cite{Vashkovskii1988} has only recently been put into the context of magnonics \cite{Tanabe2014,Hata2015}.  We show refraction and reflection of the waves and find deviations from Snell's law in optics for incidence angles larger than 25\textdegree{} with respect to the interface normal. Furthermore we can show experimentally that the spin wave amplitudes are enhanced in the vicinity of the transition region counteracting losses on lenght scales of a few micrometers.

To explain our findings, we have to incorporate the anisotropic dispersion relation for spin waves in thin films into Snell's law. Let us first consider the case of a dipolar spin wave impinging onto an arbitrary interface between two isotropic magnetic media. The continuity of the tangential component of the wave vector $k$ of any wave when experiencing reflection or when being transmitted to a different medium can be regarded as Snell's law \cite{Hecht2002,Reshetnyak2004,Kim2008}, namely 
\begin{equation}
\sin(\theta_{1})=\frac{k_{2,3}}{k_{1}}\sin(\theta_{2,3})\,,\label{eq:Snell}
\end{equation}
with $\theta_i$ the angles with respect to the interface normal. The indices 1-3 denote incoming, refracted and reflected waves, compare Fig. \ref{fig:Sketch-of-the-experiment} b) . In optics, this reduces to the well known Snell's law for refracted waves, where $k_{1,2}$ can be substituted by the respective refractive indices due to isotropic and linear dispersion relations in most materials. For the same reasons, it simply follows $\theta_{1}=\theta_{3}$ for a reflected wave since it remains in the same medium.
In contrast, the wave vector of spin waves in thin films depends on the angle $\varphi$ between the propagation direction with respect to the direction of the externally applied field $H$. This follows directly from the dispersion relation \cite{Kalinikos1986}

{\scriptsize{}
\begin{equation}
\left(\frac{\omega}{\mu_{0}\gamma}\right)^{2}=\left(H+Jk^{2}+ M-\frac{Mkd}{2}\right)\left(H+Jk^{2}+\frac{Mkd}{2}\cdot\sin^{2}\left(\varphi\right)\right)\label{eq:dispersion relation}
\end{equation}
}with $\omega$ the angular frequency of the wave, $\gamma$ the gyromagnetic ratio, $M$ the saturation magnetization, $d$ the thickness of the film, $\mu_0$ the vacuum permeability and $J=\frac{2A}{\mu_{0}M}$ with the exchange stiffness constant $A$. For the wave propagation discussed experimentally in this paper, it is safe to neglect exchange interactions (i.e. $A=0$), since we are limited to rather small wave vectors around $k=1~\mu m^{-1}$. In this range of $k$, the propagation is mainly governed by the dynamic dipolar forces originating from the precessing magnetization \cite{Kalinikos1986}.

$\varphi_{2}$ can be identified as $(\varphi_{1}+\theta_{2}-\theta_{1})$ (see Fig.~\ref{fig:Sketch-of-the-experiment}
b)) and Eq.~(\ref{eq:dispersion relation}), can be rewritten in the following form
\begin{widetext}
{
\scriptsize{}
\begin{equation}
k=\frac{\left(-\sqrt{\left(\left(H+M\right)\sin^{2}(\varphi)+H\right)^{2}-\left(2\,\sin(\varphi)\frac{\omega}{\mu_{0}\gamma}\right)^{2}}+\left(H+M\right)\sin^{2}(\varphi)-H\right)}{d\,M\cdot\sin^{2}(\varphi)} \,.\label{eq:k of H}
\end{equation}
}
\end{widetext}

This expression for the $k$-vector can be inserted into Eq.~(\ref{eq:Snell}) to obtain Snell's law for spin waves. Besides the known material and experimental parameters, the resulting implicit equation only depends on $\theta_{1}$ and $\theta_{2}$ and can therefore be used to predict refraction angles for spin waves. Similarly, the angle of reflection can be determined by identifying $\varphi_{3}=180\text{\textdegree}-(\varphi_{1}+\theta_{3}+\theta_{1})$. Feeding the calculated angles back into Eq.~(\ref{eq:k of H}) allows calculating the wave vector amplitudes. The formalism is not limited to our experiments; it can also be used for interfaces consisting of different magnetic materials.

In the trivial case of spin waves impinging at normal incidence, i.e. $\varphi_{1}=\varphi_{2}=90\text{\textdegree}$,  onto a step interface between two media with thicknesses $d_i$,  it is straightforward to define an angle-independent relative refractive index: Since $k\propto\frac{1}{d}$, $\frac{\ensuremath{k_{2}}}{k_{1}}$ in Eq.~(\ref{eq:Snell}) reduces to $\frac{d_{1}}{d_{2}}=c$. In  the experimental case discussed below $c=2$. This case corresponds to Snell's law in optics.

Experimentally, we use time resolved scanning Kerr microscopy (TRMOKE) and micro-focused Brillouin light scattering (\textmu -BLS) to verify Snell's law for spin waves.

For the TRMOKE experiments, a  800\,nm wavelength Ti:Sapphire laser is focused to a spot of 450\,nm at normal incidence onto the sample. Upon reflection, the rotation of the polarization vector of the incident light is detected which is directly proportional to the  out-of-plane component of the magnetization. In order to reach time-resolution, the laser pulses are phase locked to the microwave excitation frequency.

Using  a $x$-, $y$-, $z$-piezo stage, the sample can be scanned enabling direct access to the characteristics of the spin waves, namely wave vector, phase and relative amplitude. Simultaneously, the reflectivity of the sample is recorded which is used to identify the thick and thin parts of the sample. Typical dimensions of the images are $40\times40\,$\textmu m. We use a step size of $300\,$nm.

\textmu -BLS measurements are performed by focusing about $5\,$mW of monochromatic light from a Diode-Pumped-Solid-State laser operating at $532\,$nm onto the sample. All features of the experimental apparatus are described in detail elsewhere \cite{Madami2012}. 
Conventional BLS measurements are only sensitive to the spin wave intensity, not to its phase. In order to measure the propagation direction of spin waves it is necessary to extract the required phase information. This can be realized with the so called phase-sensitive micro-focused BLS which relies on the interference between the inelastically scattered light and a reference beam of constant phase~\cite{Serga2006}.
Two-dimensional \textmu -BLS maps are acquired by scanning the laser spot over an area of about $2.5\times2.5$ \textmu m\texttwosuperior{} with $250\,$nm step size \cite{Supplement}.

 By sputter deposition and standard lithography techniques, we fabricate a 100~\textmu m wide ferromagnetic thin film sample out of Permalloy (Py) which features a well-defined thickness step of $\Delta z$~=~30 nm, see Fig \ref{fig:Sketch-of-the-experiment} a). Spin waves are excited in the 60~nm thick part by a shorted coplanar wave guide (CPW) deposited on top of the films. The spin waves then propagate away from the CPW in the Damon-Eshbach (DE) geometry i.e. with $\it{k}$-vector $\vec{k_1}$ perpendicular to the direction of both the CPW and the applied magnetic field.
At some distance from the CPW the spin waves reach a thickness step and are refracted into a medium with lower thickness, in the present case 30~nm. 
In thin films, a change in the thickness of the magnetic material causes a drastic change of the dispersion relation which is therefore used in the experiments to model a transition to a different medium.
In fact, in Fig. \ref{fig:Sketch-of-the-experiment} b) we can clearly observe a refracted wave with altered $\it{k}$-vector $\vec{k_2}$. Similarly, a reflected wave can also be observed in the upper part of the thick region.  The angle and $\it{k}$-vector definitions are drawn on top of the experimental data obtained by TRMOKE. 

\begin{figure}
	\begin{centering}
	\includegraphics[width=0.6\columnwidth]{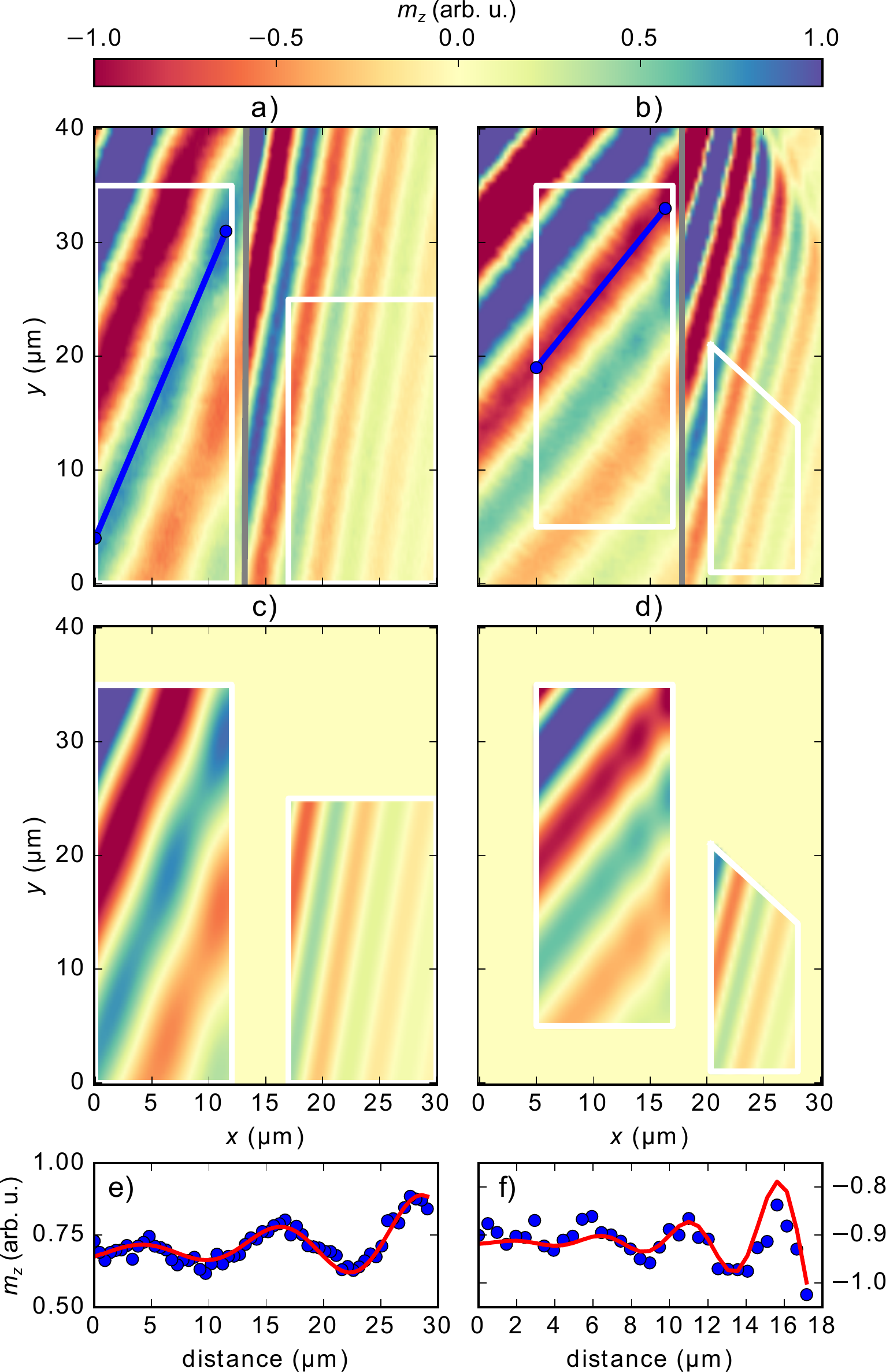} 
	\par\end{centering}
	\caption{ Experimental results for two samples. In  a) the incoming wave has an angle of $\theta_1=20^{\circ}$ with respect to the interface normal and in b) the angle is $\theta_1=40^{\circ}$.
		c) and d) show the corresponding plane wave fits. The $x$- and $y$-axis give the dimensions of the images while the color-code provides
		a scale for the dynamic magnetization component in arbitrary units.
		The gray line marks the step between thick (on the left) and thin
		(on the right) Permalloy films and the white boxes indicate the area of
		the fit. The images are recorded at a fixed frequency of $\omega=2\pi\cdot8\,$GHz
		and an external field of $\mu_{0}H=54\,$mT along the wave fronts
		of the incoming wave. The color-scale is cropped in order to enhance
		the contrast in the areas with lower signal. To emphasize the reflected waves e) and f) show line scans along the blue lines in a) and b). The blue dots are interpolated from the data; the red lines are fits extracted from c) and d), see main text. "distance" indicates the distance from the lower left to the upper right of the blue lines.}
	\label{fig:Experimental results} 
\end{figure}

In total, twelve different  samples with varying angle of incidence $\theta_{1}$ between 0\textdegree{} and 60\textdegree{} in steps of 5\textdegree{} were measured at a fixed excitation frequency of $\omega=2\pi\cdot8\,$GHz in TRMOKE experiments and $\omega=2\pi\cdot8.1\,$GHz in \textmu-BLS experiments. Examples of the raw TRMOKE data are shown in Fig.~\ref{fig:Experimental results} a) and b). In the data, we notice the incoming wave in medium 1 (left of the grey line) and the refracted wave in medium 2 (right of the grey line). When closely analyzing the dynamic magnetic contrast in medium 1, also a reflected wave can be observed. To emphasize the reflected waves we show linescans along the wavefronts of the incoming waves in Fig.~\ref{fig:Experimental results} e) and f). The crest (trough) of the incoming wave leads to a positive (negative) offset in the Kerr signal. 

As clearly seen in the images, the $\it{k}$-vector of the spin waves is significantly enhanced behind the thickness step. This means that the natural limit for short wave length spin wave generation given by the geometrical constraint of the CPW can be elegantly overcome.

 Furthermore, near the interface the signal in the thin part of the Permalloy film is substantially larger than in the thick part. This is counter-intuitive at first, since the refracted wave is induced by the incoming wave. However, the combined action of exchange and dipolar interaction leads to an increased excursion angle. To avoid dynamic magnetic charges, purely dipolar coupling would lead to a doubling of the excursion angle (since the thickness ratio of the two media is 2:1). At the same time, exchange prefers reducing the tilt angle between the precessing magnetic moments in both media. As a result, the Kerr signal increases by a factor slightly less than two. Note that also an increased in-plane shape anisotropy might contribute to the deviation from the factor of two. 
The enhancement of the amplitude is an important point and means that we can in fact boost the signal some distance from the excitation, thus counteracting natural attenuation by damping. This is a local effect, since the attenuation length in the thin part becomes shorter mainly due to the reduction of the group velocity that scales linearily with thickness. The attenuation length is further reduced since the $k$-vector increases and since the propagation direction tilts away from the Damon Eshbach geometry \cite{Kabos1994}. However, a net boost of the signal is clearly seen some micrometers from the interface.

To further analyze the experiments, we fit the data in the thin part to a 2D plane wave for the refracted waves to obtain the quantities of interest, namely wave vector amplitudes $k_{2}$ and the angles of refraction $\theta_{2}$. Additionally, amplitude, phase and attenuation length are included in this model.
The thick part is fitted with a superposition of incoming and reflected wave yielding $k_{3}$ and $\theta_{3}$. The fits are displayed in Fig.~\ref{fig:Experimental results} c) and d). The results  are also used to characterize the sample, as described in the Supplementary Material \cite{Supplement}.

For the fitting procedure, we avoid regions where the wave is disturbed by sample defects or where additional reflected waves or static demagnetizing effects near the edges of the Permalloy film alter the plane wave. This is especially important near the interface: Since components of the external magnetic field point along the interface normal, demagnetizing effects arise in the thick as well as in the thin region near the thickness step. We expect a decrease (increase) of the effective field in the thick (thin) film, as well as a tilt of the static magnetization away from the direction of the external field. Both contributions have been quantified by micromagnetic simulations \cite{Supplement}. Based on this analysis, we choose the regions for fitting where the waves can be regarded as plane waves.
More importantly, Snell's law which we probe in these experiments should still hold due to the translational symmetry along the edge.
 
\begin{figure}
	\includegraphics[width=1\columnwidth]{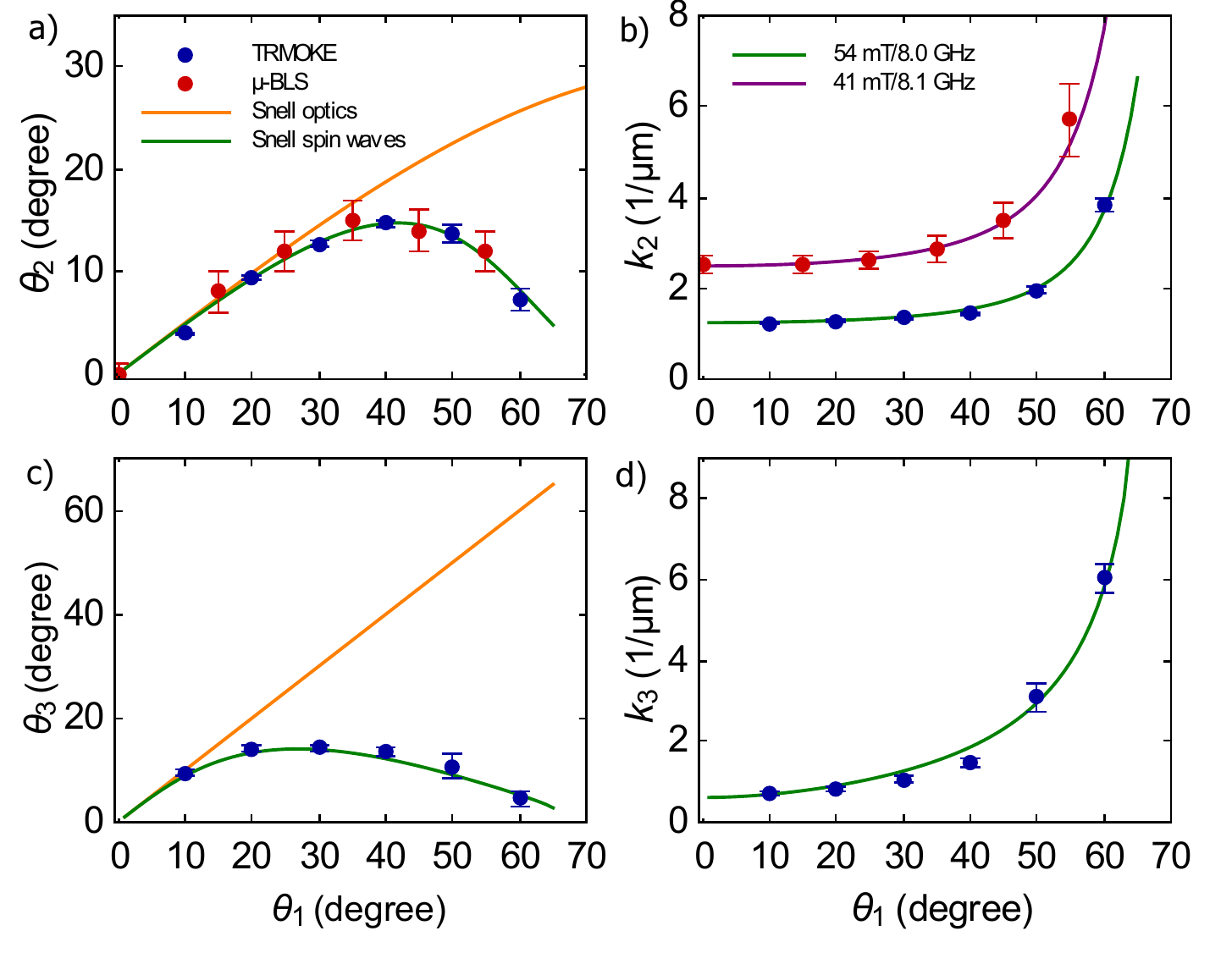}
	\caption{ a) Refracted angle $\theta_{2}$ b) refracted
		wave vector $k_{2}$ c) reflected angle $\theta_{3}$ and d) reflected wave vector $k_{3}$, all shown versus incident angle $\theta_{1}$. In all graphs, the blue dots are experimental
		values measured with TRMOKE, while red dots are measured with \textmu -BLS.
		The orange line shows Snell's law for an isotropic dispersion relation
		and the green and purple curves Snell's law for spin waves. The latter are calculated
		with the help of the anisotropic dispersion relation, eq. \eqref{eq:dispersion relation}, reflecting the different experimental conditions: The \textmu -BLS data are measured at an external field of $\mu_{0}H=41\,$mT and an excitation frequency of 8.1\,GHz, while TRMOKE data was recorded at an external field of $\mu_{0}H=54\,$mT and an excitation frequency of 8.0\,GHz. The purple curve is not shown in a), since it overlaps with the green curve. The errors are the result from least square fitting.}
	\label{fig:results of the fittinig} 
\end{figure}

The results extracted from the data are presented in Fig.~\ref{fig:results of the fittinig}. They are compared to the expectations for Snell's law in optics (orange lines), i.e. for a wave propagating in an isotropic medium. A significant deviation from Snell's law in optics is observed for incidence angles  $\theta_1 \textgreater 25^{\circ}$ in the case of refraction and $\theta_1 \textgreater 10^{\circ}$ in the case of reflection. 
One of the important results that we conclude from our experiments is that the wave vector can be very efficiently enhanced for incidence angles $\theta_1 \textgreater 25^{\circ}$. We observe that while the refracted angle starts decreasing again for $\theta_1 \textgreater 40^{\circ}$,  $k_{2}$ keeps increasing due to the anisotropic dispersion relation (in the case of reflection a decrease is observed for $\theta_1 \textgreater 25^{\circ}$) . 
Essentially, to match the condition of Snell's law and the dispersion relation at the same time, the $k$-vector needs to increase considerably for dipolar spin waves: on an iso-frequency curve, Damon Eshbach spin waves have the lowest $k$-vector.
This allows reducing the magnon wavelength efficiently. In contrast, in an isotropic system --- where the wave vector is solely determined by the refractive index which is generally not angle dependent --- it would stay constant.

One should note, that the results depend crucially on the orientation of the external magnetic field (which is aligned parallel to the antenna in all measurements),
while its magnitude is negligible for the angular dependence of the refracted wave. In contrast, the wave vector amplitude is influenced substantially by the magnitude of the external field.
This can be observed in Fig.~\ref{fig:results of the fittinig} a) and b). In the  \textmu-BLS experiments we use $\mu_{0}H=41\,$mT as external magnetic field at a frequency of 8.1\,GHz while in TRMOKE we use $\mu_{0}H=54\,$mT at a frequency of 8.0\,GHz. Since increasing the external magnetic field shifts the dispersion relation upwards, we detect a $k$-vector smaller by about a factor of two in the TRMOKE experiments (the slight frequency difference is negligible). Note that surprisingly the refracted angles remain unaffected.


We conclude that Snell's law for spin waves in the dipolar regime can be predicted with high accuracy. Our experiments can be fully reproduced by incorporating the anisotropic dispersion relation.  
We observe efficient spin wave steering due to the step interface while at the same time the wave length of the spin waves can be reduced. In the vicinity of the interface a signal boost is observed that we attribute to dynamic dipolar coupling. Our findings should be important in the field of magnonics where efficient spin wave steering remains a serious problem to be solved. For example, it can be envisaged that a series of stepped interfaces results in an increased refracted angle while at the same time short wave length spin waves can be generated.
Note that Snell's law in the form presented here, should also hold for hetero-interfaces composed of different magnetic materials. In this case the material parameters (e.g. saturation magnetization and gyro-magnetic ratio) of the different regions have to be inserted in Eq.~(\ref{eq:k of H}). 

We gratefully acknowledge funding from the following sources: JSPS KAKENHI Grant Numbers 15H05702, 25103003, 26103002, 26220711, 26870300, 26870304, Collaborative Research Program of the Institute for Chemical Research, Kyoto University, Deutsche Forschungsgemeinschaft via SFB 689. M. M. and G. G. thank the MIUR under PRIN Project No. 2010ECA8P3 "DyNanoMag". K.T. acknowledges a Grant-in-Aid for Young Scientists (B) (No.15K17436).


\end{document}